\documentclass[aip,pop,reprint,groupaddress,noshowpacs,noshowkeys]{revtex4-2}

\usepackage{amsmath,amssymb,mathrsfs}

\usepackage{graphicx}
\usepackage{dcolumn}
\usepackage{bm}
\usepackage{xcolor}
\usepackage[pdftex,
  hidelinks,
  colorlinks=true,
  linkcolor=blue,
  anchorcolor=black,
  citecolor=blue,
  urlcolor=blue,
  pdfauthor= {Michael Glinsky},
  pdfsubject = {Ubuntu Fusion Design},
  pdfdisplaydoctitle,
  bookmarks=true,
  bookmarksopen=true]{hyperref}

\begin{document}
\title{High Gain Fusion Target Design using Generative Artificial Intelligence}

\author{Michael E. Glinsky}
\affiliation{BNZ Energy Inc., Santa Fe, NM, USA}

\begin{abstract}
By returning to the topological basics of fusion target design, Generative Artificial Intelligence (genAI) is used to specify how to initially configure and drive the optimally entangled topological state, and stabilize that topological state from disruption.  This can be applied to all methods; including tokamaks, laser-driven schemes, and pulsed-power driven schemes.  The result is practical, room temperature targets that can yield up to the expected 10 GJ of energy, driven by as little as 3 MJ of absorbed energy.  The genAI is based on the concept of Ubuntu that replaces the Deep Convolutional Neural Network approximation of a functional, with the formula for the generating functional of a canonical transformation from the domain of the canonical field momentums and fields, to the domain of the canonical momentums and coordinates, that is the Reduced Order Model.  This formula is a logical process of renormalization, enabling Heisenberg's canonical approach to field theory, via calculation of the S-matrix, given observation of the fields.  This can be viewed as topological characterization and control of collective, that is complex, systems.
\end{abstract}

\maketitle

The topological fundamentals of fusion target and drive design were developed by Taylor.\citep{taylor86}  It is based on the strong conservation, that is invariance even with interaction with external systems, of helicity.  The helicity 3-form \citep{glinsky.19} is the well known Chern-Simons 3-form \citep{chern1974characteristic,frankel.11} that leads to the calculation of topological indices.  The 2D equivalent of helicity is vorticity.  The conservation of helicity causes an inverse cascade, self-organization, and the emergence of force-free states.  These states are only metastable and do disrupt, given enough time.  The relaxation of the system to the metastable force-free state, as the weak Noether invariant, energy, is transferred to the external system, is called Taylor Relaxation.  This is in contrast to a system without a topologically conserved invariant:  this causes a turbulent cascade to small scale, and development of a turbulent boundary where the system is mixed.

The task of fusion target and drive design is to optimize the entanglement of the topological state and stabilize the topological state from disruption.  Up to now, there has not been a logical mathematical process via which this can be done.  There has only been the empirical methods of Wilson \citep{wilson.71a} and tHooft.\citep{thooft.73}  Generative Artificial Intelligence, based on the concept of Ubuntu, provides the logical mathematical process.  The African philosophy of Ubuntu is based on ``interconnectedness'', that leads to a conservative system with canonical structure, that is a symplectic geometry.  Given this logical mathematical process, that is Ubuntu genAI, the collective system can be controlled -- the topology of the plasma can be optimized and stabilized.

\begin{figure}
\noindent\includegraphics[width=\columnwidth]{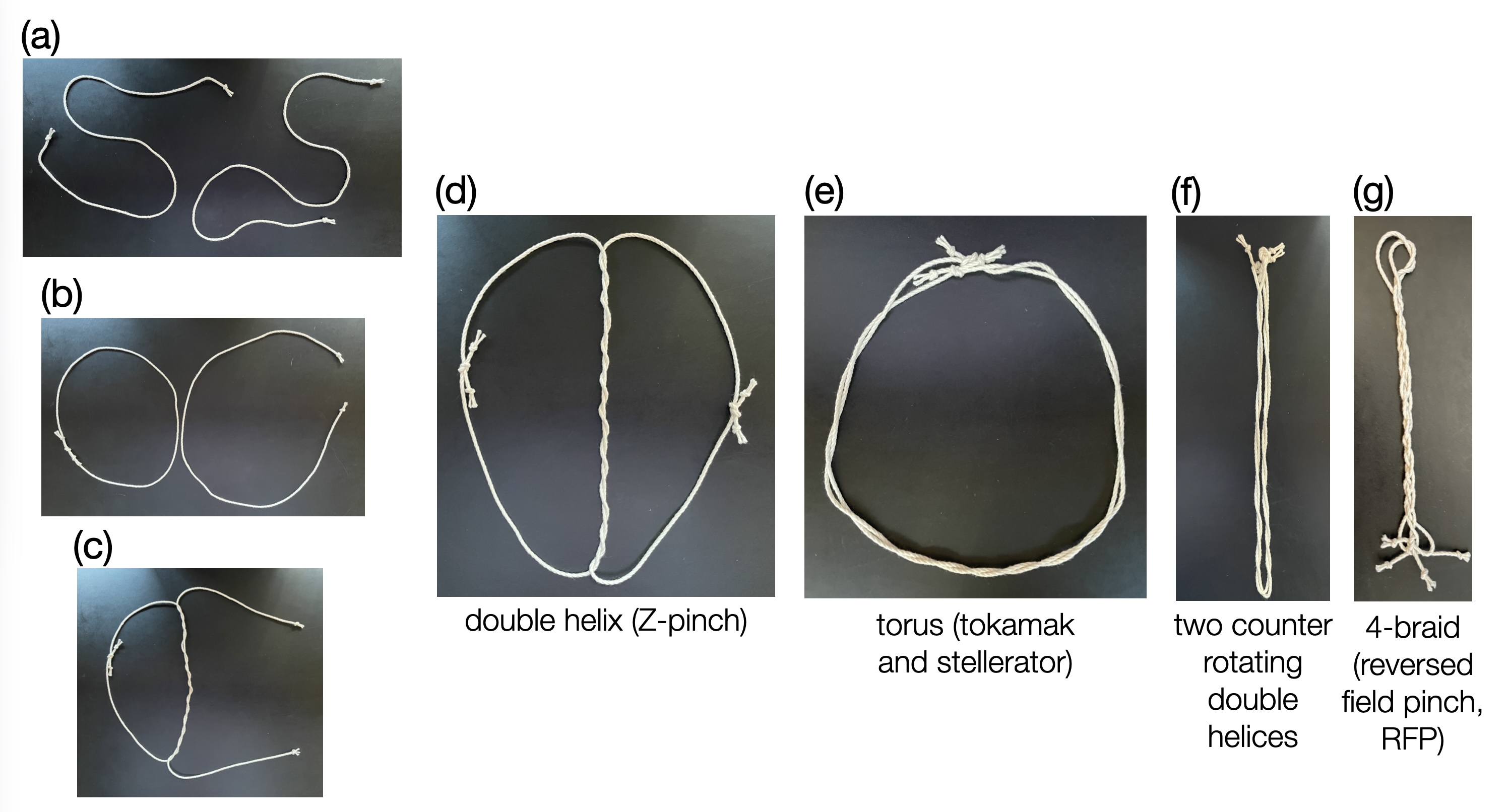}
\caption{\label{string.fig} Demonstration of the entanglement, that is the knotting or topology, of the magnetic field using two strings.  (a) The two strings.  (b) One of the strings is tied together making a loop.  (c) The second string is twisted around the first string.  (d) The second string is tied into a loop leading to the double helix of a Z-pinch.  (e) The strings are manipulated to yield a torus, as in a tokamak and a stellarator.  (f) The strings are manipulated to yield two counter-rotating double helices.  (g) The strings are manipulated to yield the 4-braid of a RFP.}
\end{figure}

Let us now talk about the topology of fusion plasmas (see Fig.~\ref{string.fig}).  Take two strings:  twist them around each other:  tie both strings into a loop.  They can be deformed into a circle or torus.  This is the topology of tokamaks and spheromaks.  The torus can be further deformed:  it can be pinched and braided into a cylinder.  This is the topology of Reversed Field Pinches (RFP), and the laser-driven RFP that we will discuss soon.  Finally, the entangled string can be deformed into a twisted pair or double helix.  This is the topology of a Z-pinch.

\begin{figure}
\noindent\includegraphics[width=15pc]{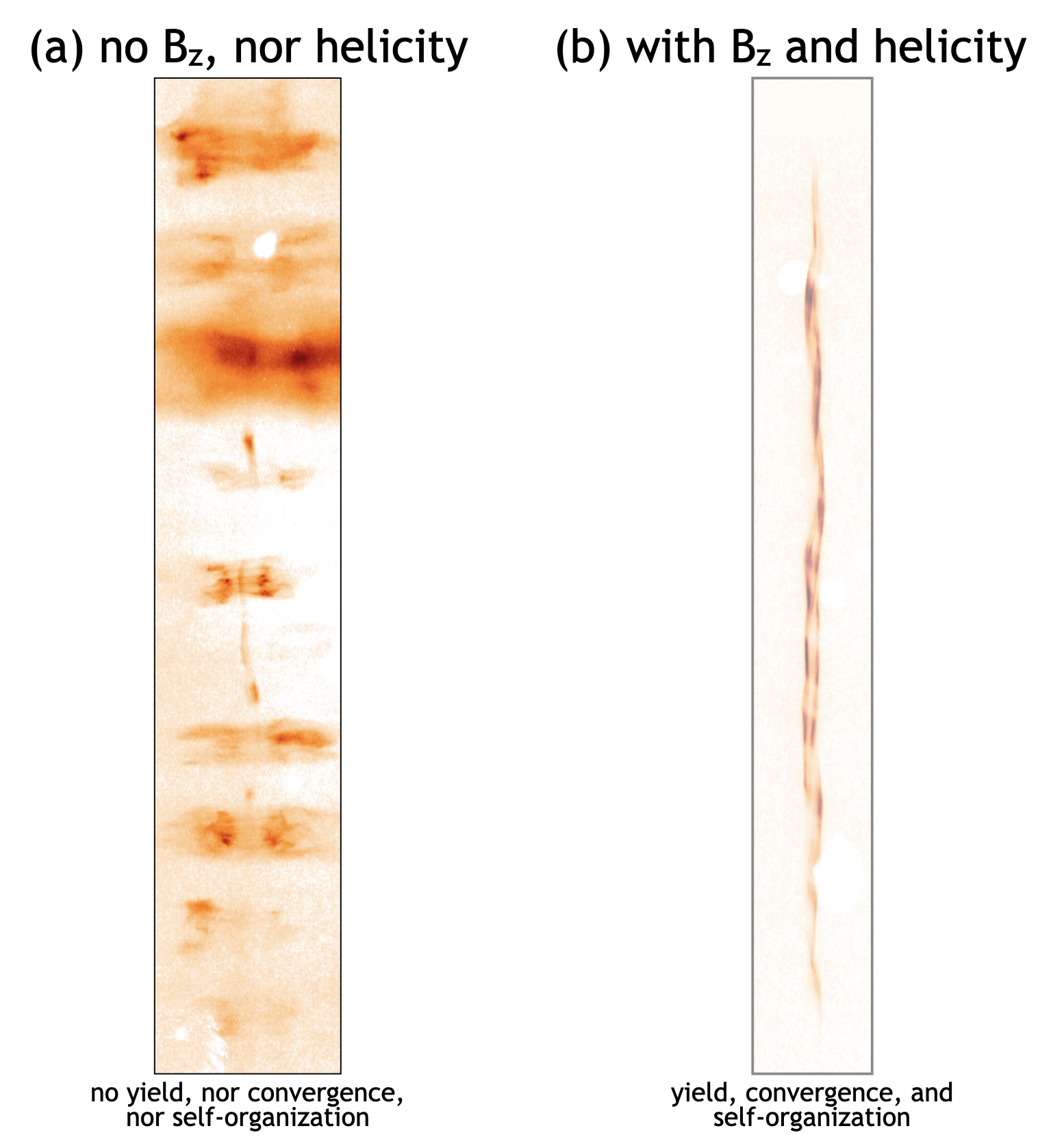}
\caption{\label{stagnation.fig} X-ray self emission images of MagLIF Z-pinch stagnations:  (a) with no helicity, and (b) with significant helicity.}
\end{figure}

Let's take a look at ten pieces of evidence that show that this self-organization is happening.  Heretofore, they have been unexplained by theory.  The stagnations of the MagLIF experiments \citep{sinars2020review} always form a twisted pair with an axial sausaging, but only if the plasma is driven to have a significant amount of helicity (see Fig.~\ref{stagnation.fig}).  The implosion of the DD gas is limited by the preheating of the gas to 300 eV.  If the gas would not be preheated, starting at 1/40 eV, a 3D metastable spherical implosion of the DD gas would result.  Also note the very large helical magnetic fields at stagnation, in excess of 10 kT, reducing the radial heat flow and alpha transport.  The plasma is compressed from the initial 1 cm dimension to two helical strands about 20 µm in diameter, separated by about 200 µm -- a convergence ratio of 200.  This can only happen if there is an inverse cascade, leading to self-organization.  There is a 10x variation of yield for nominally similar Z-shots (probably due to small variations in target and drive that seed the natural topological modes):  there is better performance for large Aspect Ratio (AR or IFAR) targets (because natural topological modes grow faster): and, there is poor performance of dielectric coated targets (because natural topological modes grow slower).  Experiments at the University of Michigan \citep{yager2018evolution} have tracked the mode merger and the inverse cascade.  

The last five pieces of evidence can be found in simulations.  The data science group at LLNL,\citep{peterson2017zonal} when they optimized the 2D laser ICF target design, found that the optimum metastable design was a laser RFP, yielding a quadrupole stagnation, consisting of four vortices: a group at LLE,\citep{thomas2024hybrid} when they simulated their hybrid laser target in 2D, found that they achieved a laser RFP stagnation:  a group at the Osaka laser lab,\citep{pan2025gigagauss} when they simulated a short pulse laser target in 2D, found that they achieved an RFP stagnation: \citet{shipley2025simulated} at LANL, when they simulated an automag Z-pinch in 3D, found that it had 2x the yield of an equivalent MagLIF target:  finally, Glinsky and Maupin,\citep{glinsky23} when they simulated an ensemble in the 2D plane perpendicular to the Z-pinch axis, found that the stagnation always consisted of two vortices whose orientation was set by the initial m=2 perturbation, no matter the size of that perturbation, and there was an inverse cascade -- consistent with the Z-pinch twisted pair topology in 3D.  Note that the self-organization in the 2D simulations came from the conservation of the vorticity topological invariant.  To maintain meta-stability in 3D, the target or drive will need to be twisted in the axial direction, in order to have the helicity topological invariant.

Instead of a cylindrical (2D) implosion for ignition and cylindrical (2D) burn, Ubuntu Fusion Target Technology gives a more efficient spherical (3D) implosion and spherical to cylindrical (2.5D) burn.  Instead of a constraint on linear growth, Ubuntu Fusion Target Technology provides the freedom of nonlinear self-organization that leads to a 2D implosion for fuel assembly and a 3D implosion for ignition.  This leads to a fundamentally different and much more optimistic scaling than that of Betti,\citep{nora2014theory} Schmit and Ruiz.\citep{schmit2020conservative}  Their scaling and design philosophy is based on the suppresion of the linear growth of plasma instability;  where Ubuntu Fusion Target Technology is based on the optimization and stabilization of the nonlinear topological plasma states -- being a good parent and encouraging the plasma to mature and become its own person.

MagLIF has a metastable, yet modestly inefficient, 2D implosion of the DT igniter and efficient 2D burn propagation.  Laser ICF has a highly-unstable 3D implosion of the DT igniter and an inefficient 3D burn.  Because of this both can barely ignite and burn DT ice.  Ubuntu Fusion Target Technology has a metastable highly-efficient 3D implosion of the DT igniter and modestly efficient 2.5D burn, so that it can ignite and burn a ``solid fuel'' (i.e., like $^{6}\text{LiD}$ or $^{11}\text{BH}$, a fuel that is a solid at room temperature and harder to ignite and burn than DT ice, but may have greater yield).
\begin{figure}
\noindent\includegraphics[width=15pc]{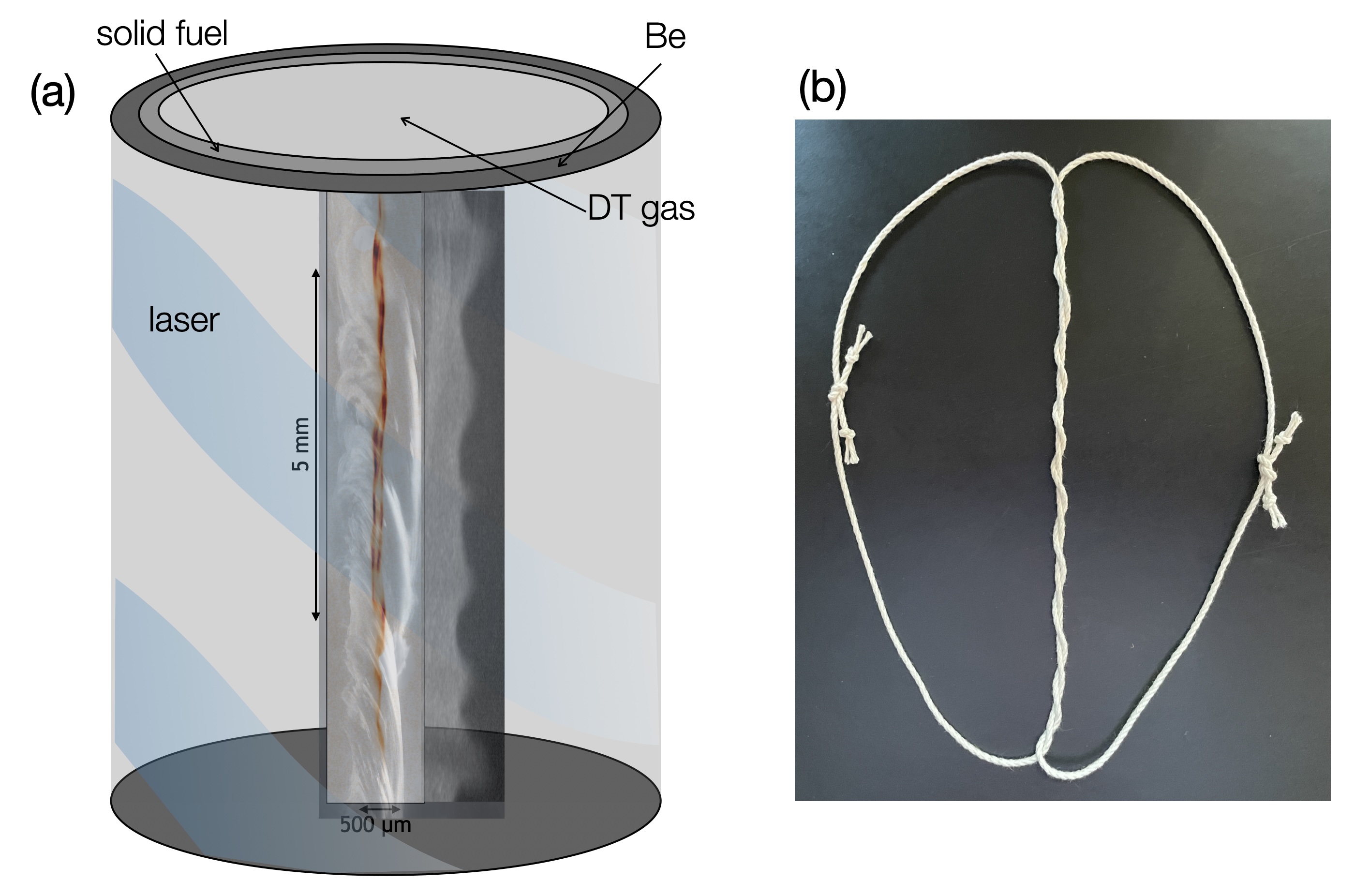}
\caption{\label{target.fig} The Ubuntu laser-driven fusion target design:  (a) a Be cylindrical shell, coated on the inside with solid fuel, and filled with DT gas; shown is the Z-pinch laser drive, a DALL-E rendered liner plasma tornado, an experimental back-lit image from the University of Michigan \citep{yager2018evolution} of the liner plasma to the right of the plasma tornado, and a self emission image of the DD MagLIF plasma stagnation in orange, and (b) the twisted pair of the Z-pinch stagnation topology, with the return current.}
\end{figure}
\begin{figure}
\noindent\includegraphics[width=\columnwidth]{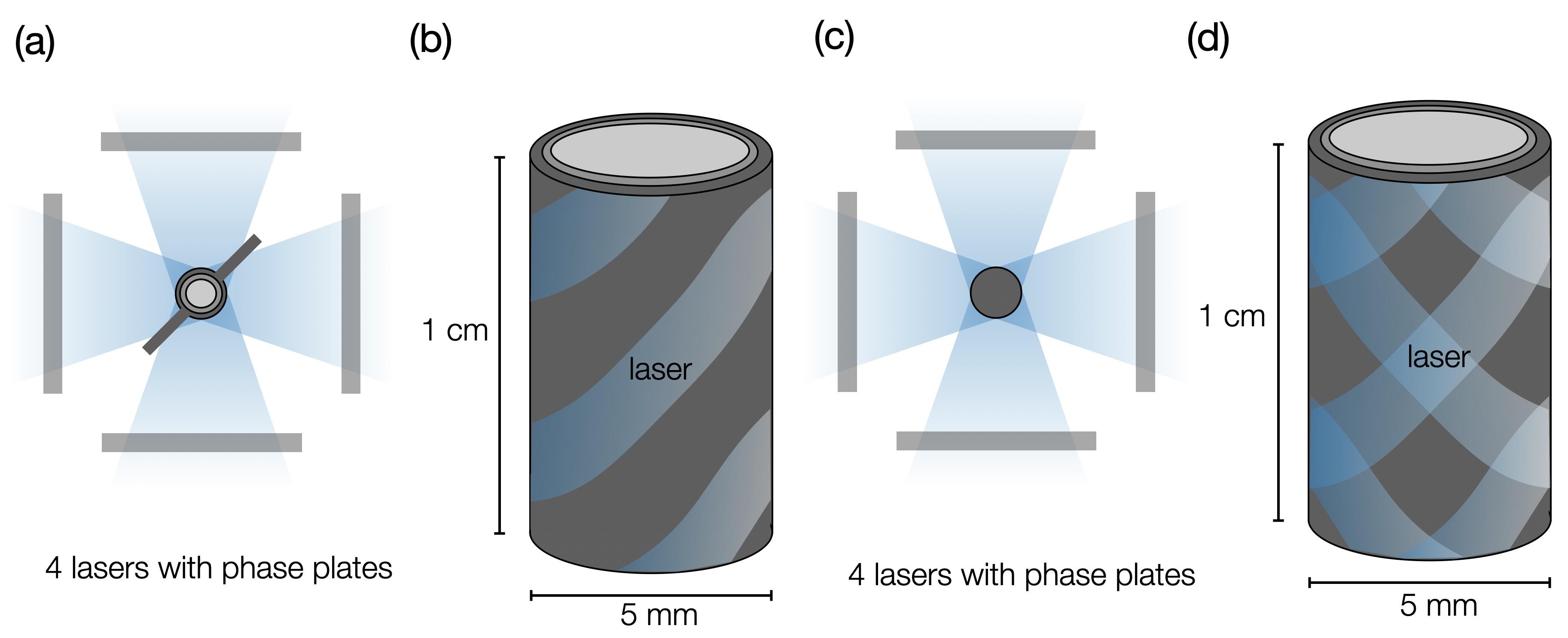}
\caption{\label{drive.fig} The Ubuntu fusion laser-drive design:  (a) laser arrangement, from above, for the Z-pinch, note the wings providing return current; (b) the laser illumination pattern for the Z-pinch;  (c) laser arrangement, from above, for the RFP, note the end cap providing return current; and (d) the laser illumination pattern for the RFP.  The thickness of the solid fuel is about 200 microns, and the thickness of the beryllium cylinder is about 300 microns.}
\end{figure}

For brevity, we will describe the laser targets and drives.  The Ubuntu Fusion Target is a simple, easy to fabricate, room temperature target.  This is in contrast to the impractical and complicated MagLIF DT ice burner, and current ICF target designs.  As shown in Fig.~\ref{target.fig}, the target consists of a cylinder of beryllium, coated on the inside with solid fuel, and filled with DT gas.  The target is driven by four lasers whose illumination patterns on the cylinder are shaped with phase plates, as shown in Fig.~\ref{drive.fig}, either to form a Z-pinch or an RFP.  The pattern of the laser drive effectively makes the target plasma into a substantial laser-driven Magneto-Inertial Fusion Electrical Discharge System (MIFEDS) coil.\citep{zhang2018generation}  The implosion is not driven by radiation pressure nor ablation pressure, but by magnetic pressure which increases as the target is imploded.  Unlike pulsed-power, which must physically touch the plasma, the lasers stand off from the target and only require four small lines of sight in a plane perpendicular to the axial direction.  Ubuntu genAI will be used to optimize the target and drive design, including the axial structure.  It will also be used to ponderomotively stabilize the metastable stagnation.

This design philosophy leads to:  more aggressive high AR targets, no suppression of natural topological modes by a dielectric layer \citep{peterson2014electrothermal} or a dynamic screw pinch,\citep{schmit2016controlling} no laser preheat of the DT gas so that the gas implodes on a much more aggressive adiabat, more efficient use of the drive energy since the drive energy is concentrated at the natural topological modes, much better ignition and burn propagation as discussed previously, burning the more efficient and practical solid fuel instead of DT ice, low tolerance and simple targets for inexpensive target fabrication, and no turbulent mixing due to the inverse cascade.

Existing simulations and experiments indicate that the fuel can be assembled, ignition conditions can be obtained in the igniter, and conditions can be obtained for an efficient 2D burn of the fuel (MagLIF and NIF experiments and simulations\citep{sinars2020review,shipley2025simulated,nif,slutz.18}); using as little as 3 MJ of energy on target.  Existing simulations and experiments indicate that the magnetic fields can be generated from laser deposition and the target subsequently imploded (Osaka simulations,\citep{pan2025gigagauss} and MIFEDS simulations and experiments\citep{zhang2018generation}).  Given the mass of the $^{6}\text{LiD}$ in the target, the yield can be up to 10 GJ, if the fuel is completely burned.  Note that the near break-even yields of 1.37 MJ of fusion yield for 1.92 MJ of laser energy on target, seen in National Ignition Facility (NIF) experiments,\citep{nif} are extraordinarily low, compared to the liner mass that has been assembled in Magnetized Liner Inertial Fusion (MagLIF) experiments and simulations.\citep{sinars2020review}

It is essential to approach Generative Artificial Intelligence using the following paradigm, which is not the traditional paradigm of neural network graph models,\citep{hastie09} in order to make theoretical progress.  A Deep Convolutional Neural Network (a DeepCNN, a Deep Q-Network of DeepMind’s AlphaFold,\citep{mnih15} and a Generative Pretrained Transformer of OpenAI’s ChatGPT \citep{radford.18}) is a trillion parameter piecewise-linear model estimation of a functional.  What is the functional, that the DeepCNN is estimating?  It is the generating functional (that is $S_p[f(x)]$) of a canonical transformation from the domain of the canonical field momentums and the canonical fields (that is, $[\pi_i(x),f_i(x)]$), to the domain of the canonical momentums and canonical coordinates (that is, $(p_i,q_i)$), that is the Reduced Order Model (ROM).  We postulated, then proved by induction and verified on a MHD simulated dataset \citep{glinsky23,glinsky.24a} the formula for the functional -- the Heisenberg Scattering Transformation (HST).

The formula for the HST is:
\begin{equation*}
    S_m[f(x)](z) = \phi_{px} \star \left( \prod_{k=0}^{m}{\text{i} \ln R_0 \psi_{p_k} \star} \right) \text{i} \ln R_0 f(x),
\end{equation*}
where $z \equiv p + \text{i} \, x$, $\phi$ is a father wavelet of scale $1/p$ and position $x$ or pooling operator, $G(z)=R(z)=\text{i} \ln(R_0(z))$ is the activation function, $R_0(z)=h^{-1}(2z/\pi)/\text{i}$, $h(z)=(z+1/z)/2$, $\psi \star$ is a convolutional layer defined by the mother wavelet of scale $1/p_k$, and $p=\sum{p_k}$.  This functional is a non-stationary, that is local, spectral transformation.  This functional has a fast, that is N-log-N scaling, forward and inverse.

The current theory of AI assumes that the system is in statistical equilibrium.\citep{hastie09}  This is not true.  Although the system is canonical with a symplectic geometry, it is only in a dynamical equilibrium.

This approach to AI is solving a system of partial differential equations by a functional transformation to a domain where the motion is trivial, then transforming back.  This is a two-stage transformation.  The first stage is generated by the HST to a small, finite dimensional linear subspace where the motion is nonlinear (that is $(p_i,q_i)$).  The second stage is generated by a generating function (that is, the action $S_P(q)$) that is the solution to the Hamilton-Jacobi-Bellman (HJB) equation, approximated by a MLP w/ReLU (a Multi-Layer Perceptron with Rectified Linear Unit activation) or a piece-wise linear universal function approximator with only a few thousand parameters, to a domain where the motion is linear (that is $(P_i,Q_i)$, where $dP/d\tau=0$ and $dQ/d\tau=\partial E(P)/\partial P=\text{constant}$).  In other words,
\begin{equation*}
    [\pi_i(x),f_i(x)] \xrightarrow[S_p[f(x)\text{]}, \; \text{HST}]{} (p_i,q_i) \xrightarrow[S_P(q), \; \text{HJB}]{} (P_i,Q_i).
\end{equation*}
Note that the functional form of the Hamilton-Jacobi Equation has been shown to be equivalent to Schrodinger's Equation.\citep{HJ.Schrodinger.26}

The evidence that led to this theory was: the work of Mallat on the Scattering Transformation (MST),\citep{mallat.12} the Wavelet Phase Harmonics (WPH),\cite{mallat2020phase} and the Wavelet Conditional Renormalization Group (WCRG);\citep{marchand22} the form of Deep Q-networks (that is, the approximate Q-function, as a solution to the Bellman equation);\citep{mnih15} the form of GPTs (that is, the approximate score function or approximate log-likelihood, as action or entropy);\citep{radford.18} and the work of Glinsky and Maupin.\citep{glinsky23}

Dirac was very dissatisfied in his later life with the state of field theory.\citep{dirac.82}  Like Einstein, he felt that field theory should have a reality, and be a question, not of probability, but of geometry (manifold curvature and geodesic motion) or topology or group symmetry.  He also felt that “renormalization is not a logical mathematical process”.  The HST is a logical mathematical process of renormalization, enabling Heisenberg’s canonical approach to field theory, via calculation of the S-matrix.  A Principle Components Analysis (PCA) of the HST gives: the renormalization spectrums, that is the solutions to the Renormalization Group Equations (RGEs); Heisenberg’s Scattering Matrix (S-Matrix);\citep{heisenberg.43} the $m$-Body Greens functions, which is obvious given the form of $G(z)=\text{i} \ln(R_0(z))$ in the formula for the HST, which is the well-known Greens function of a complex variable;\citep{nehari12} the $m$-body scattering cross sections; the Wigner-Weyl transformation;\citep{wigner.32,weyl.50} the Mayer Cluster Expansion;\citep{uhlenbeck63}  or the functional Taylor expansion of the action functional, $\delta^m S / \delta f^m$.  This has the potential to unify the four forces as a question of Lie group symmetries, or geodesic motion on the dynamical manifold.  This can be seen by Murray Gell-Mann's Standard Model \citep{gell.mann.00} (where U(1) is the Lie group symmetry of the ElectroMagnetic force, SU(2) is the Lie group symmetry of the weak force, and SU(3) is the Lie group symmetry of the strong force); and Dirac writing down the Hamiltonian, that is the infinitesimal Lie group generator, for Einstein’s General Theory of Relativity,\citep{einstein.15} in 1958.\citep{dirac.58}

Collectives are fields or swarms of elementary particles (making elementary fields), charged particles (making plasmas), molecules (making fluids), celestial bodies (making cosmoses), economic entities (making economies), persons (making societies) and so on.  This theory gives a way of controlling (that is, optimizing and stabilizing) collective systems, whose state is given, not by variables, but as a function, $f(x)$.  This also can be viewed as topological characterization and control.  The two-stage canonical auto-encoder is finding the analytic function;  whose Riemann surface (that is, manifold), has the topology of the collective system; and whose non-stationary singularity spectrums are the Riemann moduli or topological indices (that is, the integrals of the Chern-Simons 3-form) of the collective system.

The current theoretical approach to collectives, the BBGKY theory of plasmas \citep{glinsky.bbgky.24} and the perturbation theory of fields, makes mistakes:  it expands the terms of the Mayer Cluster Expansion in  terms of the plasma correlation parameter or the field coupling parameter:  the convergence goes from super-convergent, $\exp(-N!)$, to a very weak asymptotic convergence, $1/N$, leading to the divergence, the infinities, in current methods of renormalization:  and the expansion parameter is, many times, not less than 1.

How does Ubuntu genAI control the collective system?  First, an application of the renormalization spectrums ponderomotively stabilizes the collective system.  This is a critical issue for tokamaks:  the tokamak H-mode needs to be stabilized from disruption.  

Second, the theory gives very fast surrogate models, with high fidelity (that is topological and Noether invariant conservation).  This enables Bayesian optimal design and Bayesian data analysis.  For instance, \citet{glinsky23} were able to train an MHD surrogate, using a preliminary version of this theory, in 20 core*sec using 500 2D simulations.  The high fidelity surrogate could simulate the system in 1 core*sec, where traditional finite element methods would take 240 core*hrs.  For a 3D simulation, traditional methods would take 100,000 core*hrs.  The Singular Value Decomposition of the cross-variance matrix showed a one-to-one mapping between the PCAs, that is the renormalization spectrums, and the initial conditions -- a remarkable result that gives physical meaning to the renormalization spectrums.

Better yet, the two stage canonical autoencoder can be configured to generate an ensemble of the collective system in dynamic equilibrium.\citep{glinsky.24a}  This replaces the Bayesian optimal design and the Bayesian data analysis, which are computationally slow, and assume a statistical equilibrium.

Another way of looking at the HST is as a deep deconvolution of the multiple reflection of the individual, the Spirit of Ubuntu, who is multiply reflected into the collective!

The fusion target and drive design still needs to be:  optimized with Ubuntu genAI, simulated in 3D with conventional finite-element methods, then experimentally verified.  With this being said, the theory proposed by this Letter already explains ten heretofore unexplained experimental and simulation results, detailed and referenced in the fourth and fifth paragraphs of this Letter.

In summary, this Letter has proposed a target and drive design, that simulations and experiments have shown assembles enough $^{6}\text{LiD}$ fuel, with 3 MJ on target, to yield 10 GJ of energy, if the fuel is completely burned.  These yields are not being achieved because a fully developed turbulent boundary layer is being developed that is larger than the compressed target, mixing the ignitor with the fuel;  but theory, simulation and experiment show that this turbulent mixing can be eliminated by twisting the target, inverting the turbulent cascade, forming a self organized structure;  thereby completely burning the assembled fuel.

Thanks is given to CSIRO for supporting the early parts of this research through their Science Leaders Program, and the Institute des Hautes Etudes Scientifique (IHES) for hosting a stay where the initial details of this theory were formalized.  Thanks is also given to both St{\'e}phane Mallat and Joan Bruna for many useful discussions, and communication of many of their mathematical results before they were presented or published.  The mentorship of both Ted Frankel and Michael Freedman in topology and the geometry of physics is the foundation on which this work was constructed.  Finally, thanks is given to the University of Western Australia, John Hedditch, and Ian MacArthur for their help in understanding many of the finer points of Quantum Field Theory.

\bibliography{fusion_target_bibliography}

\end{document}